%% file: difftestpaper.tex
\documentclass[]{article} 

\usepackage{paralist}
\usepackage{booktabs}
\usepackage{epsf}
\usepackage[T1]{fontenc}
\usepackage{listings}
\usepackage{graphicx}

\usepackage[usenames,dvipsnames]{color}

\usepackage{url}
\usepackage{eulervm}

\usepackage{microtype}
\usepackage[hidelinks]{hyperref}
\usepackage[square,sort&compress]{natbib}
\usepackage{amssymb}
\usepackage[scaled]{beramono}
\renewcommand\cite[1]{\citep{#1}}

\input{aspectjlistings}

\input{common}

\newcommand\text[1]{\textsf{#1}}
\lstdefinelanguage{kconfig}{
  keywords={config, if, endif, default, bool, depends, on, select, choice, endchoice, boolean, tristate, prompt},
  keywordstyle=\bfseries,
  ndkeywords={m, n, y},
  ndkeywordstyle=\color{blue}\bfseries,
  identifierstyle=\color{black},
  sensitive=false,
  comment=[l]{//},
  morecomment=[s]{/*}{*/},
  commentstyle=\color{purple}\ttfamily,
  stringstyle=\color{red}\ttfamily,
  morestring=[b]',
  morestring=[b]",
}
\begin{document}

\title{Differential Testing for Variational Analyses: \\ Experience from Developing KConfigReader}

\author{Christian K{\"a}stner\\
Carnegie Mellon University
}
\date{}

\maketitle
\begin{abstract}
Differential testing to solve the oracle problem has been applied in many scenarios where multiple supposedly equivalent implementations exist, such as multiple implementations of a C compiler. If the multiple systems disagree on the output for a given test input, we have likely discovered a bug without every having to specify what the expected output is. 
Research on variational analyses (or variability-aware or family-based analyses) can benefit from similar ideas. The goal of most variational analyses is to perform an analysis, such as type checking or model checking, over a large number of configurations much faster than an existing traditional analysis could by analyzing each configuration separately. Variational analyses are very suitable for differential testing, since the existence nonvariational analysis can provide the oracle for test cases that would otherwise be tedious or difficult to write. 
In this experience paper, I report how differential testing has helped in developing KConfigReader, a tool for translating the Linux kernel's kconfig model into a propositional formula. Differential testing allows us to quickly build a large test base and incorporate external tests that avoided many regressions during development and made KConfigReader likely the most precise kconfig extraction tool available.
\end{abstract}

\section{Introduction}

Over the years, my collaborators and I have built many different analysis tools for 
highly-configurable software systems, including CIDE~\cite{KA:ASE08}, TypeChef~\cite{KGREOB:OOPSLA11,KOE:OOPSLA12,LRKADL:ESECFSE13},
and Varex*~\cite{NKN:ICSE14,MWKTS:ASE16}. Those tools were designed to perform complicated analyses
at scale on real languages, be it Java, C, or PHP. Getting them right required to be precise about
language semantics and the variability mechanisms that we analyzed. We spent
a lot of time testing our implementations on small and large examples. But over time
our quality assurance strategy changed, adopting more and more the idea of \emph{differential testing} for configurable systems.
In this paper, I share my experience of systematically using differential testing
while building \emph{KConfigReader} in 2014, which was pivotal for me and
led me to strongly advocate differential testing in all subsequent projects. 
I hope this experience might convince others to adopt similar quality assurance strategies
in their work as well.

KConfigReader is a tool to read the Linux kernel's variability model (\emph{kconfig} files)
and translate it into a propositional formula for automated reasoning with SAT solvers. 
It is similar to prior tools in LVAT~\cite{SLBWC:VaMoS10,SB:kconfig10} and Undertaker~\cite{TSSL:FOSD09,TLSS:EUROSYS11}
which did all the heavy lifting in understanding the \emph{kconfig} semantics and 
pioneering such translation.
I needed such a tool for analyzing the Linux kernel for type and linker errors with \emph{TypeChef},
to report only errors that were not already excluded by the variability model.
Although we originally used LVAT in TypeChef, it had known bugs and limitations and
was no longer maintained, and Undertaker's infrastructure was not designed for the accuracy needed in TypeChef.
In the spring of 2014, I (naively) decided to write my own tool, aiming for \emph{maximum precision} 
in the translation from \emph{kconfig} into propositional formula.

KConfigReader was the first project in which I consequently used differential testing
from the very beginning. For a test case with 10 boolean options, there are only
1024 potential configurations---a number small enough to perform some computation
on each configuration separately.
Using small test cases with less than 10 options, I ran the 
original \emph{kconfig} program on all possible configurations of each test to get a ground truth about which
configurations were valid, without having to specify the oracle myself. This allowed
me to quickly write test cases even for corner cases, to explore unusual constructs, and to build
a regression test suite that would catch mistakes introduced when addressing other
corner cases. Differential testing was essential to implementing accurate transformations
and helped make KConfigReader likely the most accurate tool of its kind, as recently 
also confirmed in an independent study by \citet{EKS:GPCE15}.

In this paper, I want to share my experience with differential testing in the context
of variational analyses. 
I learned that differential testing is an ideal match for testing variational analyses,
that setting up a differential-testing infrastructure is usually fairly simple,
and that once set up it drastically simplifies and encourages writing tests.
I have used differential testing on several other projects
since, and even used it to create tests for earlier projects. While the idea is not 
new, I think it is severely underappreciated in our community and I believe that many
other researchers and practitioners could benefit from it.

\section{Differential Testing for Variational Analyses}

\begin{figure}
	\centering
	\includegraphics[scale=.4,trim={4.5cm 6.5cm 4.5cm 5.2cm},clip]{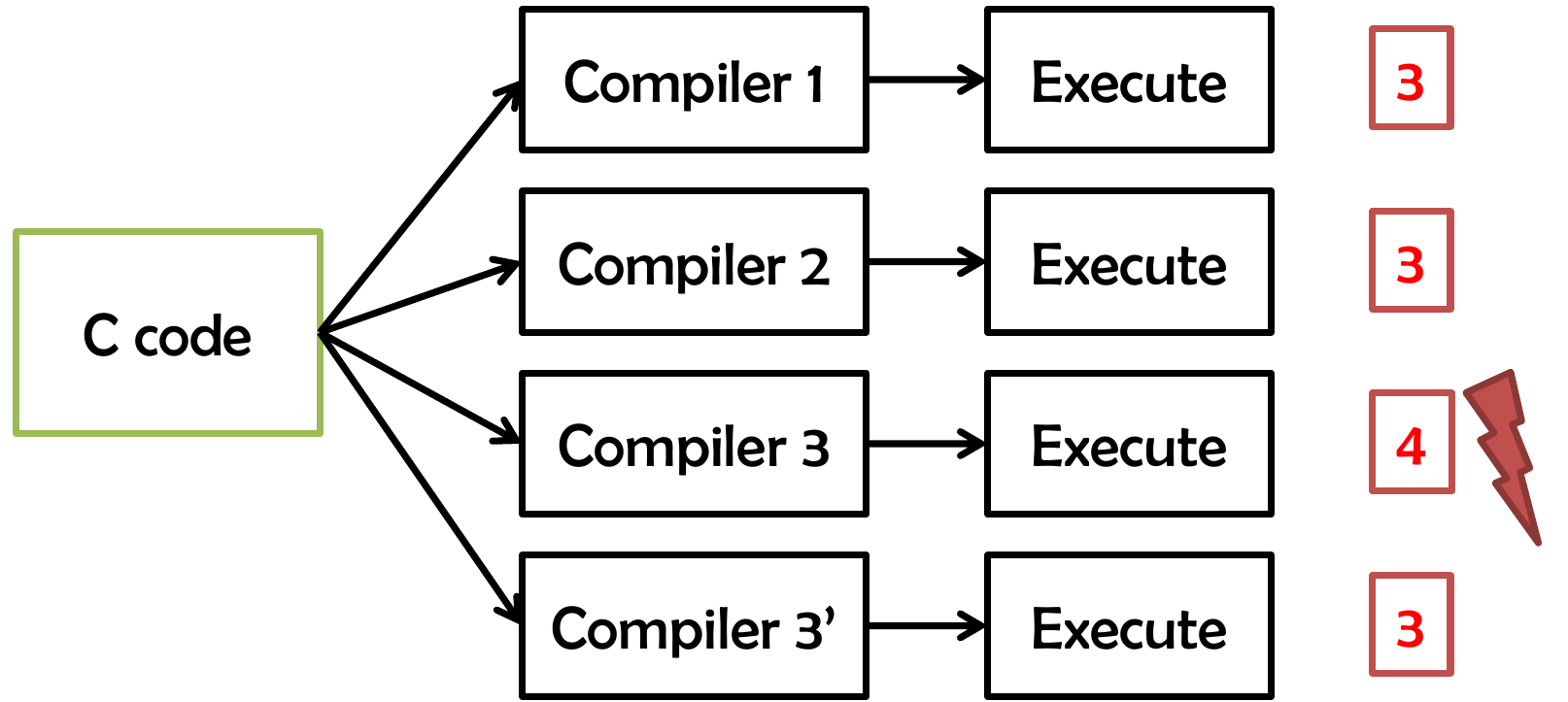}
	\caption{Differential testing of compilers: Detecting compiler bugs by looking for disagreement of the results of running a program compiled by different compilers.}
	\label{fig:csmith}
\end{figure}

My first conscientious exposure to differential testing was
Yang et al.'s PLDI 2011 paper ``Finding and Understanding Bugs in C Compilers''
on \emph{CSmith}~\cite{YCER:PLDI11}. The authors developed an infrastructure
to generate random C code snippets to test the optimization phase in compilers.
Since it is difficult to predict the expected correct output for a randomly generated
code fragment, they used the simple idea of compiling the code with multiple compilers.
If the programs compiled with different compilers from the same source code behave
differently, there is likely a bug in at least one compiler. In Figure~\ref{fig:csmith},
this setup is illustrated.
With their implementation
\emph{CSmith}, the authors found and reported hundreds of compiler bugs.

In a nutshell, differential testing solves the oracle problem of testing:
Given a test case, how do we know whether the test execution was successful or not.
In unit testing, developers often provide the oracle manually in terms of writing
assert statements. This is labor intensive and easy to get wrong, writing assertions
that are too weak to detect subtle incorrect behavior or that are wrong and encode
incorrect exceptions are not uncommon.
With differential testing, we use an existing implementation as the oracle. If the
new implementation disagrees with the existing one on a test input, we fail the test,
without ever having to write any manual assertions. Differential testing makes it
much easier to write test inputs, since we do not need to provide the expected results.
Therefore it is even possible to generate test inputs mechanically.

Differential testing has been used in many contexts for many years~\cite{W:DTJ98,GHJ:ICSE07,LS:ICTCS06,PMFB:ISSTA10,YCER:PLDI11}.
While very useful, it is only applicable in contexts where a reference
implementation exists (or multiple implementations to allow voting to
decide which implementation contains the bug).
Interestingly, essentially all \emph{variational analyses} that I have been working
on fulfill this criterion and are thus amendable to differential testing.

In a nutshell, variational analyses (or variability-aware or family-based analyses) work as follows: They perform some
sort of analysis on entire configuration spaces of highly-configurable systems.
During the analyses they exploit the similarities among different configurations
and avoid the redundancies of analyzing each configuration separately in a brute-force
fashion. Since the configuration space tends to grow exponentially with the number
of configuration options, a brute force analysis is infeasible for but the smallest
systems; in contrast, variational analyses have been shown to scale to huge configuration spaces 
with hundreds or even thousands of configuration options  (having configuration
spaces with more possible configurations than there are atoms in the universe).
For example, variational type checking encodes differences among configurations
compactly and checks types for all configurations through a SAT encoding~\cite{TBKC:GPCE07,CP:GPCE06,KATS:TOSEM12,LRKADL:ESECFSE13}.
Variational analyses have been explored for parsing, type checking, data-flow analysis,
model checking, and others---for an overview see a recent survey~\cite{TAKSS:CSUR14}.

\begin{figure}
	\centering
	\includegraphics[scale=.4,trim={2cm .5cm 1cm 3cm},clip]{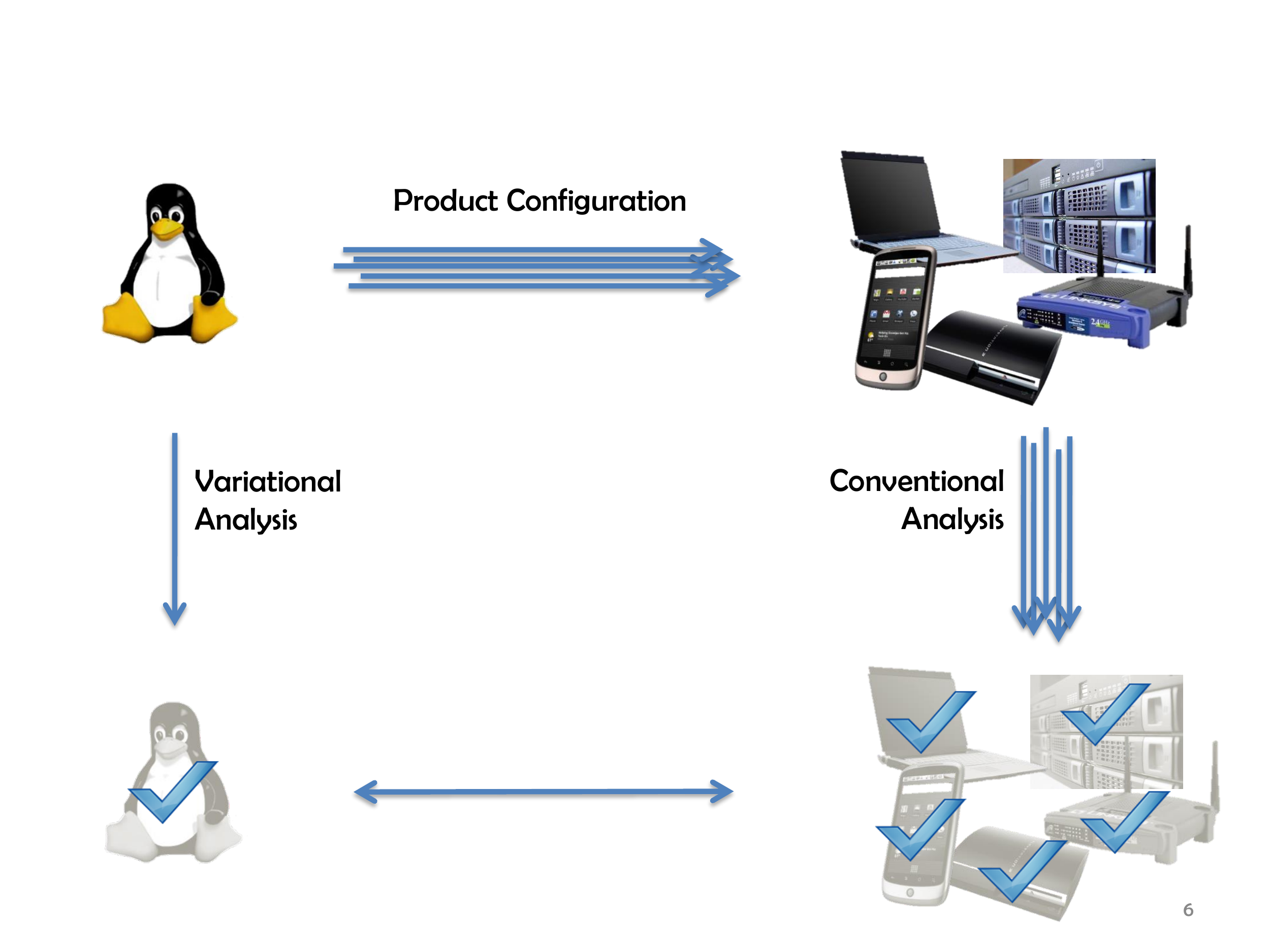}
	\caption{Variational analysis and brute-force execution of traditional analysis yielding equivalent results.}
	\label{fig:commuting}
\end{figure}

Differential testing is a good fit for variational analyses because most variational 
analyses aim to be \emph{sound and complete with regard to applying
an existing tool in a brute-force manner}, as illustrated in Figure~\ref{fig:commuting}. 
Generally, variational analyses aim to find all issues
that could be found with existing approaches and no additional ones, just much 
faster. 
Therefore, an existing nonvariational reference implementation (the
`conventional analysis' arrow in Fig.~\ref{fig:commuting})
can provide the oracle when executed separately for each configuration.
For example, TypeChef aims to efficiently parse unpreprocessed C code with arbitrary
preprocessor directives while only rejecting input for those configurations for which a 
traditional C parser would fail after preprocessing. 

For differential testing, we simply exploit the equivalence between (a) a
variational analysis and (b) applying a traditional analysis in a brute-force
fashion. 
Where traditionally differential analysis would check two or more programs
performing the same task (e.g., compilers in Fig.~\ref{fig:csmith}), we 
compare the result of the variational analysis with the
outcome of the existing tool for individual configurations.  For small
configuration spaces (say up to 10 boolean options), we can apply the
traditional analysis to every single configuration and thus establish ground
truth for the entire configuration space; for large configuration spaces we
can still use this to test select configurations.

While the idea seems obvious in retrospect and I hope that others have
used this form of testing for their variational analyses, I am not aware of
earlier systematic applications. This paper intends to spread the idea.

\section{Developing KConfigReader}

KConfigReader parses the variability model of the Linux kernel, specified  in
the domain-specific language of the kernel's kconfig tool, and translates it
into a large propositional formula. The formula should describe exactly those
configurations that can be configured with  kconfig. The formula can be used
for many purposes; for example, TypeChef uses it to avoid false positives by
issuing a type error only when it can occur in at least one valid
configuration.

Honestly, I severely underestimated the involved complexity of the kconfig
configurator.
There are several challenges in this translation. First, the semantics
of the kconfig language are not well described and the kconfig implementation
is the only reference (though core aspects have been formalized by \citet{SB:kconfig10}). 
Second, the language includes features beyond propositional logic, including
3-value logic for options and constraints over numeric options.
Third, kconfig's configuration language is not fully declarative but 
follows an imperative configuration mechanism in which
the order of selecting options can matter. Forth, kconfig changed slightly throughout
the Linux kernel's evolution, requiring updates of the translation tool.

Although prior tools existed for this translation (see introduction), they did
not model kconfig's semantics accurately enough to be used confidently in the
context of the analyses we do with TypeChef. TypeChef reported several errors
with constraints that should have been excluded by the variability model, but
due to inaccurate translation we spent much time trying to understand bugs
that could not be reproduced, because we could not create corresponding
configurations with kconfig.  My goal with KConfigReader was to provide a
solution that was accurate and that I could understand and maintain as
kconfig evolves.

\paragraph{Crash course in kconfig.}

To understand how KConfigReader works and how it can be tested, let us
briefly introduce how kconfig works.
The Linux kernel's kconfig language provides a mechanism to describe
configuration spaces in a textual form in kconfig files. A kconfig file describes options and their dependencies,
as well as additional information about options, including textual explanations.
The kconfig tools read a kconfig file and present interactive
configuration dialogs to the user, who can select or deselect options
as long as constraints are fulfilled (multiple graphical and textual frontends exist).
The selected configuration is then written into a local file (\emph{.config})
that is used by the build system during the compilation process to
include or exclude files or code fragments within files.

In Figure~\ref{fig:test1} (left), we illustrate a small example of a configuration
file, representing three Boolean configuration options \emph{A}, \emph{B}, and
\emph{NOPROMT} grouped together in a \emph{choice} to indicate that exactly
one of them can be selected. Each option that can be shown in an interactive
dialog is represented by a textual prompt and an optional help text.
Defaults and dependencies can be described with corresponding declarations.

The kconfig language has a number of nuances and quirks. Options without a
prompt are invisible and can never be changed by a user, but are automatically computed based
on dependencies and defaults.
Beyond the shown boolean options, also \emph{tristate} options, string options, and
numeric options are supported. For tristate options, users can configure
\emph{`y'} (selected), \emph{`m'} (compile as module), or \emph{`n'} (deselected),
in which `compile as module' is a specific feature in the Linux build infrastructure
to build a code unit as separate module rather than linking it statically into the kernel.
Numeric and string options are typically used for token substitution in the 
source code (as C preprocessor macros). Boolean, tristate, and numeric options
can be used in complex constraints, such as ``\lstinline[language=kconfig].depends on (IA64 || X86) && USB_BUS='m' && CPU>3.''.
The interaction among these features is often nontrivial~\cite{SB:kconfig10,EKS:GPCE15}.

\begin{figure}
\centering
\begin{minipage}{.4\linewidth}
\begin{lstlisting}[language=kconfig]
choice
		prompt "choice prompt"

config A
    boolean "A prompt"

config B
    boolean "B prompt"
    default n

config NOPROMPT
    boolean
    default y

endchoice
\end{lstlisting}
\end{minipage}
\hspace{2em}
	\begin{tabular}{ll}\toprule
		$\{\}$ &X \\$\{\text{A}\}$ &X \\$\{\text{B}\}$ &X \\$\{\text{NOPROMPT}\}$ &X \\$\{\text{A}, \text{B}\}$ &X \\$\{\text{A}, \text{NOPROMPT}\}$  & $\checkmark$
		\\$\{\text{B}, \text{NOPROMPT}\}$ & $\checkmark$
		\\$\{\text{A}, \text{B}, \text{NOPROMPT}\}$ & X
		\\\bottomrule
	\end{tabular}

\caption{Example test file and corresponding  ground truth produced by kconfig.}
\label{fig:test1}
\end{figure}

\begin{figure}
\centering
	\includegraphics[scale=.4,trim={0 .5cm 1cm 1cm},clip]{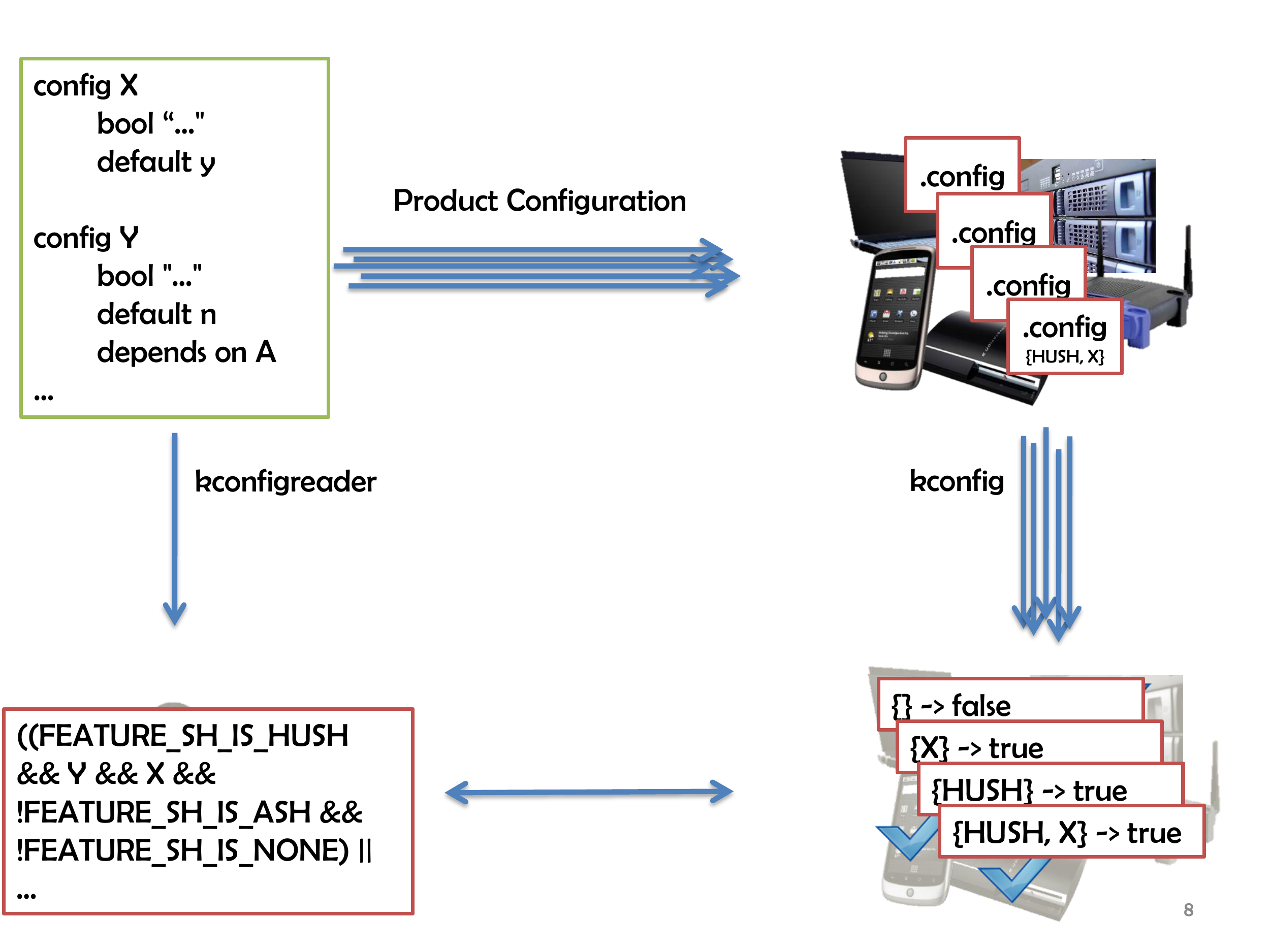}
	\caption{KConfigReader tested against a brute-force execution of kconfig itself.}
	\label{fig:commuting2}
\end{figure}

\paragraph{Differential testing.}
Knowing about the difficult corner cases and complicated semantics of kconfig
from prior debugging experience, I decided to use differential testing
during the development of KConfigReader from the very beginning 
(in fact, the very first commit was mostly test infrastructure together
with a very simple translation).
Differential testing in KConfigReader works as follows, illustrated in Figure~\ref{fig:commuting2}:
\begin{itemize}
	\item A test file, such as the one shown in Figure~\ref{fig:test1} (left), tests
	some kconfig snippet with up to 10 configuration options.
	\item KConfigReader translates the model into a propositional formula---in this example $\text{NOPROMPT} \wedge ((\text{A} \wedge \neg\text{B})\vee(\neg\text{A} \wedge \text{B}))$.
	Variables in this formula refer to configuration options; given an assignment for a specific configuration, the formula should evaluate to \emph{true} if and only if the configuration is valid.
	Details of the translation, and how some options may be encoded with multiple variables, are described in the appendix.
	\item The test infrastructure collects all used options and builds a set of all configurations (combinatorial explosion; up to 1024 for 10 boolean options)---in this example the eight configurations shown in Figure~\ref{fig:test1} (right) for the 3 boolean options in our example.
	\item For each configuration, the test infrastructure creates a \emph{.config} file in kconfig's format and runs kconfig's \emph{conf} command-line util as an oracle. Without human interaction, the \emph{conf} util reads a configuration and repairs it if it is invalid. Thus, if the \emph{.config} file remains unmodified the configuration was valid, otherwise it was invalid. Using the \emph{conf} util as oracle, we can automatically build a truth table indicating which configurations are valid as shown in Figure~\ref{fig:test1} (right)---it is not necessary to manually specify the expected result for the test snippet.
	\item If the extracted propositional formula disagrees with any result from the brute force execution of \emph{conf}, that is, if evaluating the formula for any configuration yields a  result different from running \emph{conf} on that configuration, the test for the given snippet fails---in our example, the test passes. Alternatively, it would be possible to translate the test results (truth table) into a disjunction ($(\text{A} \wedge \neg\text{B}\wedge\text{NOPROMPT})\vee(\neg\text{A} \wedge \text{B}\wedge\text{NOPROMPT})$) and check equivalence of that ground-truth formula against KConfigReader's extracted formula with a SAT solver---the former strategy makes it easier to provide meaningful error messages though.
\end{itemize}

While this differential testing strategy can obviously not be used to test the
entire kernel's variability model or even any of its nontrivial submodels, it
allows us to write tests cheaply, without effort to also provide the correct
answer. For example, in the test in Figure~\ref{fig:test1}, the expected
answer is not entirely obvious---option $\text{NOPROMPT}$ is not visible
during the configuration and does not count toward the choice, but is always
enabled---one could have easily provided the expected test output incorrectly.

Throughout the development of KConfigReader, 
I essentially performed test-driven development, writing test cases
for a particular feature of the kconfig language first, observing what result kconfig
would produce for all configurations and then implementing that feature.
The interactions of choices, dependencies, tristate options, and
invisible options turned out to be particularly tricky (see appendix) and having
an incrementally growing test suite of kconfig models testing various
corner cases provided immensely useful for regression testing.
In many cases, the implementation of one feature (say dependency propagation
among items within a choice) seemed plausible, but subtly broke other features.
It turned out that translating certain constructs accurately required sophisticated 
encodings in propositional logic
producing fairly big formulas.
Due to all those complicated cases, the translation code ended up 
far more complex than I initially hoped, and I am pretty sure that I could
not have developed and maintained it without the test suite.

\paragraph{Generated and stolen test cases.}

Whereas the original work on differential testing always combined 
an existing implementation as oracle with random test case generation,
I wrote most of my test cases manually.

I experimented with generating test cases, but found it difficult
to write a generator that would write interesting models that would
contain complicated corner cases of various options combined with
dependencies, choices, and other features. Writing more intelligent
generators would be an interesting project in its own right, but seemed
way too time consuming for the benefit expected.

A much more interesting benefit was that I could easily steal test cases
from other projects. 
Other projects also independently thought about possible corner cases
and wrote several test cases; merging them was beneficial
to ensure that I did not miss any behavior covered by tests in existing tools.
Specifically, I included the test cases from the existing tools LVAT and 
Undertaker in my test suite. 
Even though they manually specified
the expected formulas that should represent the result of analyzing 
test files, I could simply copy their 
kconfig files without using their oracles or having to adopt anything else of 
their testing infrastructure. In fact, I discovered that some of
those test cases had incorrect oracles in the original project and
some tests could not even be executed with the \emph{conf} tool because they
were not fully conforming to kconfig's syntax.

After \citet{EKS:GPCE15} performed their comparative study of 
kconfig translation tools, I could also include
their test cases, including one failing one that revealed an error I could fix.%
\footnote{Two other issues reported in that paper are corner cases
explicitly not supported by the translation discussed as limitations in the appendix.}

At this point KConfigReader's test suite contains 86 manually written test cases,
21 test from LVAT, 32 test from undertaker, 15 test from El-Sharkawy et al., and 100~randomly generated tests.
All these tests including their automatically derived oracles could
also easily be reused for testing other tools in the kconfig context.

\section{Further Differentially-Tested \\Variational Analyses}

Colleagues and I have used differential testing for a number of other projects
on variational analyses. The idea is always the same: find a non-variational
application that can serve as an oracle when executing tests in a brute-force fashion.

\paragraph{Variational Execution.}
We used differential testing systematically in much of the development of our different
attempts of \emph{variational execution}~\cite{NKN:ICSE14,MWKTS:ASE16}.\footnote{\url{https://github.com/ckaestne/Varex2},
\url{https://github.com/chupanw/vbc},
\url{https://github.com/meinicke/VarexJ}} The idea of variational execution
is to execute a test case with concrete values over a large configuration
space, while preserving as much sharing as possible to avoid redundant 
computations. Again, as illustrated in Figure~\ref{fig:commuting}, the result
of variational execution (i.e., failing or passing test cases and even values of variables at any point in the execution) should be equivalent to that of executing each
configuration separately. 

To test our variational execution engines (modified interpreters and a 
mechanism to rewrite bytecode), we used existing execution
engines as oracle. For example, for our PHP implementation, we simply use
the original PHP interpreter without modifications as the oracle for testing
our modified interpreter.
We developed a lightweight testing infrastructure, such that we could
write PHP code with special tokens (`@A', `@B', and so forth) that represent configuration options, such as the following:
\begin{lstlisting}[language=PHP, ndkeywords={@A,@B},ndkeywordstyle=\color{blue}\bfseries,]
$a = @A + 1; 
if (@B) { $a = 5; }
while ($a < 10) { echo $a; $a++; }
\end{lstlisting}
The variational interpreter understands these tokens as conditional 
values representing both \emph{true} and \emph{false} in different configurations,
whereas for the brute-force execution they are replaced by `1' or `0' in 
a preprocessing step for each configuration. The existing interpreter then
simply executes the plain PHP code of each configuration (with substituted configuration tokens)
and records the printed output from \emph{echo} statements as oracle for that configuration.
 In addition to writing our own tests, we use the existing test suite of the
interpreters (without variability) as additional test cases to ensure
that we do not break existing nonvariational behavior when modifying
the interpreter.

In our Java versions, we pursued a very similar strategy but pushed the
comparison even further. Instead of just comparing whether a test 
case crashes or what output it prints, we compare actual execution
traces. For the oracle, we use an instrumented but nonvariational
execution in which we track which bytecode instructions are executed
in what sequence. We subsequently compare this with the trace of
bytecode instructions executed in the variational execution, recorded
with similar instrumentation. This way, we can use arbitrary snippets
of Java code as tests and ensure that they are executed in a variational
fashion without breaking Java's semantics.

Finally, we started using the same testing infrastructure also for benchmarking.
Measuring both the execution time for the variational execution over all
configuration as well as the execution times for the brute-force execution, we
could track speedups of our tooling for evaluations and performance regression
testing.

\paragraph{TypeChef.}
Our TypeChef infrastructure should parse, type check, and further
analyze all compile-time variation of unpreprocessed C code~\cite{KGREOB:OOPSLA11,KOE:OOPSLA12,LRKADL:ESECFSE13}.
That is, the result of running TypeChef should be equivalent
with the result of running a traditional type checker on every
configuration of the same code, produced by running the C preprocessor.

There are many steps in TypeChef that could have benefited from
differential testing. Unfortunately, TypeChef was not originally developed with
differential testing; we added differential testing only later and selectively.
That is, it helped, but we also missed a number of opportunities for differential testing:
\begin{itemize}

  \item  TypeChef's lexer should produce a token stream that is equivalent
with the token stream produced by the C preprocessor on each configuration.
We initially developed TypeChef only with tests with handwritten oracles.
Even though it was easy in retrospect, we did not add a differential test infrastructure for the lexer
until very late, long after the initial releases.
The lack of better testing in early phases of the development caused many problems, not least because we built on an open-source Java
reimplementation of the C preprocessor that turned out to not fully implement
the exact semantics of the preprocessor. 
Interestingly, \emph{SuperC}~\cite{GG:PLDI12} was developed around the same time to 
solve the same problem and we eventually used differential testing to compare
the output of the TypeChef lexer with that of the SuperC lexer (where admittedly
essentially all differences can be attributed to bugs in TypeChef).

  \item TypeChef's parser should produce the same
AST from unpreprocessed C code as from parsing the preprocessed file
in each configuration. Since we developed our own AST, a direct comparison
with ASTs produced by \emph{gcc} or other parsers would be not
be particularly useful, but even comparing the result of our own parser
on preprocessed and unpreprocessed C code could potentially reveal
inconsistencies and bugs.

  \item TypeChef's type checker should produce the same error messages   for
type errors in unpreprocessed C code as a standard compiler   would produce
when run on each preprocessed configuration separately.   Unfortunately, we
never used differential testing for this purpose;    it could probably have
saved a lot of effort and could have produced   a more accurate type checker.

  \item Also other infrastructure built on TypeChef, including data-flow analysis~\cite{LRKADL:ESECFSE13},
pointer analysis~\cite{FMKPA:SPLC16}, and refactoring engines~\cite{LJGAL:ICSE15,MRGAKFCF:TSE17}
could likely have benefited from differential testing. Even if no
nonvariational program could produce the same output directly on preprocessed
C code as oracle, we could have compared whether the variational execution on
unpreprocessed C code was the same as the execution of the same tool on
preprocessed C code.

  \item For evaluating correctness of the refactoring engines \emph{Morpheus}~\cite{LJGAL:ICSE15} and 
  \emph{Colligens}~\cite{MRGAKFCF:TSE17} for unpreprocessed code, 
  also a variant of differential testing was used. Instead of comparing a variational
  with a nonvariational refactoring engine---for example, ensuring that
  refactoring on unpreprocessed C code yields exactly the same result as 
  refactoring preprocessed code in each configuration---they exploit a different
  mechanism as oracle that is common in testing refactoring engines: 
  The code before and after a refactoring should behave the same.
  For example, \citet{LJGAL:ICSE15} tested that a refactored program still
  compiles in the same configurations it compiled before the refactoring and
  that the test suite yields exactly the same results after refactoring
  than before---both properties are tested on tests with small enough configurations spaces 
  by compiling the program and executing the test separately for each configuration
  both before and after the refactoring. \citet{MRGAKFCF:TSE17} pursued
  a similar strategy with generated tests. Note the difference how KConfigReader was tested 
  (cf.\ Fig.~\ref{fig:commuting2}): For KConfigReader, a single execution of a variational 
  analysis was compared against  the brute-force execution of an existing tool, which established the oracle, whereas the refactoring 
  engines were tested by comparing a brute-force execution of the original code
  against a brute-force execution of the refactored code. This highlights once more the power of
  differential testing when suitable oracles can be identified.

\end{itemize}

Overall, our experience shows that developing variational analyses is
difficult, but that it can often benefit from differential testing, not just
for KConfigReader. In cases where we did not use differential testing, as in
most of TypeChef, I regret it. Where we used it, as in variational execution,
I believe it helped us significantly. Once the traditional analysis that can
be used to produce the oracle is identified, the cost of  setting up
variational testing is typically relatively low. The ease of writing tests
that can even catch minor deviations without having to provide the oracles is
liberating.

\section{Conclusion}

Already when developing early versions of the TypeChef lexer and parser in 2010 
my colleagues Sebastian Erdweg and Tillman Rendel suggested testing TypeChef against
a brute force approach. Unfortunately, I did not listen back then, discouraged
by the additional effort for writing what I now recognize to be a differential testing infrastructure.
It was over three years later, when I first experienced the benefits of differential testing during the KConfigReader
development. Looking back, I wish, I had recognized the benefits earlier and I believe
it would have helped us significantly with making TypeChef more accurate and more compatible
with existing tools. With the KConfigReader experience, I am now convinced and have used
differential testing in several projects. I also have convinced collaborators to adopt differential testing
as well. This paper is my attempt to proselytize the rest of the community.

\appendix

\section{KConfigReader: Propositional Encoding and Limitations}

In this appendix, I will provide some further details of how
KConfigReader works internally and what its limitations are. 
This is not intended to be a complete description of how to 
translate the KConfigReader semantics, but should illustrate
the concepts used in the translation and the limitations of the
current implementations. It may help to understand why the 
transformation is nontrivial and why differential testing was 
 beneficial.

In general, KConfigReader parses a kconfig file and collects all
options and choices. It then creates a, typically very long, list
of constraints expressed as propositional formulas that describe
which options can have which values and which options can not
be selected together. Most translations are local in that a single
configuration option can be translated into a small set of constraints,
but some constraints also require investigating multiple options or
choices together. The conjunction
of all produced constraints forms the propositional kconfig model. 
This model evaluates to true exactly for all assignments that
correspond to configurations that the kconfig model accepts.
For reasoning with a SAT solver, the model is translated into the
common dimacs format.

In 2016, the x86 kconfig model of the Linux kernel had about
11,000 options. KConfigReader produces about 70,000 constraints
that translate to a 14mb dimacs model with about 60,000 variables and 620,000 clauses.

\paragraph{Tristate options.}
Tristate options in kconfig represent a form of three-value logic~\cite{SB:kconfig10}
that can be freely intermixed with boolean options. KConfigReader
models a tristate option $o$ internally with three-value logic
but translates it to propositional logic using two mutually exclusive boolean variables
$o_y$ and $o_m$ that indicate when option $o$
has value `\emph{y}' or `\emph{m}' (condition $\neg o_y\wedge\neg o_m$ 
indicates that $o$ has value `\emph{n}').
The same way, every expression over tristate options can be
modeled with two boolean variables for each expression, so that
expressions can now be translated as follows:

\vspace{1em}
\noindent
\begin{tabular}{ll}
$e=\neg a$ & $e_y=\neg(a_y\vee a_m) \quad e_m=a_m$\\

$e=a\wedge b$ & $e_y=a_y\wedge b_y \quad e_m=(a_y\vee a_m)\wedge (b_y\vee b_m)\wedge\neg(a_y\wedge b_y)$\\

$e=a\vee b$ & $e_y=a_y\vee b_y\quad e_m=(a_m\vee b_m)\wedge\neg a_y\wedge\neg b_y$\\
\end{tabular}
\vspace{1em}

These translations can be uniformly applied also for combinations of boolean and tristate options, where
$b_m$ is always set to false for boolean options. As a further complication, a global \emph{MODULES}
option can control whether tristate options are handled as boolean options for the entire
configuration model.

\paragraph{Numeric and string options.}

Numeric and string options can have large or even infinite domain, such
that we do not want to model every possible value. 
Instead, we model each relevant value we find in the kconfig model with a distinct variable.
Specifically, we collect all known values of numeric options specified as defaults, in range
constraints, or in literals used in comparisons with this value. For string options, we collect
all values in defaults. 
We then model each known value with a boolean variable and create constraints to declare
them mutually exclusive.
For example, for numeric option $n$ with known values $0, 5, 100$
represented by the mutually exclusive variables $n_0$, $n_5$, and $n_{100}$.
For constraints that involve numeric values, we compute possible results for each known value and
replace the constraint with a boolean expression over the corresponding variables.
For example, the expression \emph{n<=5} is translated to a constraint $n_0\vee n_5$. 

\paragraph{Option dependencies.}
Options can declare dependencies on other options, where dependencies
are expressed as formulas over expressions, where expressions can
compare options to literals or other options.
For example, \lstinline[language=kconfig].config A boolean depends on B='n' || C='y'. can
be translated into the constraint $A_y\Rightarrow \neg (B_y\vee B_m)\vee C_y$.
Dependencies for tristate options can be encoded as well with
$o_y\Rightarrow d\_y$ and $o_m\Rightarrow d_y\vee d_m$ for option
$o$ and dependency $d$.

In addition, to dependencies declared on an option, reverse dependencies of other
options can affect an option. With a reverse dependency \lstinline.config O select P if c., option $O$
can activate option $P$ under condition $c$.
For the purpose of establishing constraints, we handle such reverse
dependencies essentially equivalently to a declared dependency on target option 
(\lstinline.config P depends on O && c.). Due to technicalities of how reverse dependencies work,
they are a known source of inaccuracies as discussed below.

\paragraph{Invisible options.}

Options without a prompt are not accessible to users during configuration, but
are still influenced by defaults and dependencies. For example, an invisible
option can be declared to be selected by default, which is changed if and only
if a dependency restricts it. To further complicate translation, whether an
option has a prompt can depend on other options and a default value can be computed from
other options as well. If an option has multiple defaults, the
first that fulfills its dependencies will be chosen.

Translation of invisible options is not trivial but works roughly like this:
A boolean option $O$ with a prompt under condition $p$ and with the default `\emph{y}' and constraint $c$
(\lstinline[language=kconfig].config O boolean "prompt"  if p default 'y' depends on c.)
will be translated to a constraint $\neg p_y \Rightarrow (c_y \Rightarrow O_y)$---that is, if there is no prompt then $O$ will always be selected
unless the constraint $c$ is not fulfilled. Nonboolean options and multiple positive
and negative defaults complicate the rules further.

\paragraph{Choice.}
Choices as in Figure~\ref{fig:test1} allow users to select exactly one of the inner options,
or multiple `\emph{m}' values for a tristate choice.
Intuitively, this is achieved by constraints that require that at least one of 
the inner options is selected and no two options are selected at the same time.
Unfortunately, in combination with tristate options and invisible options, the translation
is far from obvious. For example, options are not considered
as children of a choice when they are invisible (which may depend on
other options), but when a invisible child is selected by some constraint, also the outer
choice must be selected; tristate options within a boolean choice behave like
boolean options; a tristate choice itself can be selected as `\emph{y}' or
`\emph{m}' and behaves differently in either case, supporting either multiple
selections of inner options as `\emph{m}' or only a single option as `\emph{y}'.
The specific encodings can be found in the implementation, and many test cases
illustrate those corner cases.

\paragraph{Limitations.}

While KConfigReader is designed with the goal of accurately translating
all of kconfig, there are some limitations. Some of those limitations
account for special cases that I did not consider worth addressing, whereas
others are conceptual limitations that are difficult to overcome.
With each limitation come opportunities for further extensions.

First, the most severe inaccuracy relates to reverse dependencies (\emph{select} clauses). 
Reverse dependencies are executed whenever the corresponding option is selected
and may break other constraints in the process. In fact, they act as imperative
`\emph{if x gets selected, select y}' logic, rather than a declarative constraints 
`\emph{if x is selected, y must be selected.}' As a consequence the order of selecting
options may matter. Kconfig issues a warning whenever a reverse dependency
violates other constraints, but allows the violation nonetheless. My understanding
is that this effect is not intentionally used and even actively discouraged, but that it is rather a side effect of how kconfig
is implemented. In fact, it would be worth building an analysis on top of the models produced by 
KConfigReader to point out the situations in which a reverse dependency can
overwrite other dependencies as a potential warning for the maintainers of the kconfig
model.

Second, for numeric and string options, we collect all values that occur in defaults
and constraints, but do not model other values. While sufficient for some purposes, our 
models cannot decide validity of configurations with other values for these options,
and we do not support range expressions over variable ranges.
While string options seem rarely relevant, as they never occur in constraints, 
numeric options could be implemented using classic encodings with a variable for each of their bits
at the costs of larger constraints.
Alternatively, one could use a different target formalism and reason with SMT solvers rather
than SAT solvers.

\paragraph{Opportunities.}

I believe that accurate propositional models of kconfig produced by KConfigReader
can be valuable for a large number of purposes, beyond their current use to
check the relevance of errors produced by TypeChef.

Most prominently, propositional models could be used in a configurator that would use
the model and SAT solvers rather than the existing imperative logic
to improve the experience of users configuring their kconfig models.
While KConfigReader can provide the model, a large body of research on reasoning about feature models can provide a path for
building better configuration tools, including tools that can help users with reconfiguration (\emph{`What do I need to change to enable this option?'})
or can explain why certain options are disabled~\cite{HXC:VAMOS12,XHSC:ICSE12,BSR:IS2010,B:SPLC05}.

Furthermore, analyses can help maintainers of kconfig models to detect
inconsistencies. Classic analyses can detect dead and false optional options
or unnecessary constraints~\cite{BSR:IS2010},
or could detect refactorings~\cite{TBK:ICSE09}. Beyond that, new analyses targeted
specifically at kconfig issues are possible: For example, it would
likely be possible to develop an analysis to detect where reverse dependencies
could violate configuration constraints (see limitations above) or detect
corner cases with (tristate) choices that may lead to surprising effects
for users.

An accurate propositional model of a kconfig file can also be useful for
tool developers that analyze the Linux kernel. Beyond determining
whether individual configurations are valid (e.g., to filter warnings
produced by an analysis tool), such model can also be used to produce
valid configurations for testing, either randomly\footnote{Any valid solution produced by a SAT solver will provide a valid test 
case. Achieving true randomness 
in a highly constrained space is a research problem in itself though.} or
in combination with other sampling strategies~\cite{MKRGA:ICSE16,CDS:ISSTA07}.

Finally, there are certainly also opportunities for further optimizations. 
For example, KConfigReader produces a large propositional formula consisting of thousands of
constraints (for the Linux kernel) but maintains only weak traceability to the
origin of each constraint. For several analysis and debugging tasks it might
be worth to track specific constraints back to their sources in the kconfig files.
Similarly, precise handling of choices, invisible, and tristate options often lead to large
propositional constraints.  This seems unavoidable unless imprecise approximations are desired
for faster reasoning.
It could be worth providing both an underapproximation and an overapproximation for
situations where faster reasoning is required but false positives or false negatives
are acceptable.

\bibliography{dblp2_short} 

\end{document}

%% file: aspectjlistings.tex
\lstdefinelanguage{aspectj}[]{Java}{
	morekeywords={pointcut,call,aspect,advice,
		execution,around,abstract,extends,returning,within,args,
		target,after,before,withincode,super,privileged,
		declare,error,precedence,cflow,cflowbelow,proceed,parents}, 
	sensitive=f,
	deletendkeywords={Long}
}[keywords,comments,strings]
\lstdefinelanguage{guidsl} {
	morekeywords={implies,or,and,not,\%\%},
  moredelim =[l][\slshape]{::},
  morecomment=[l]{//},
  breaklines,breakatwhitespace
}
\lstdefinelanguage{Scala}%
  {morekeywords={abstract,case,catch,class,def,% 
    do,else,extends,false,final,finally,%
    for,if,implicit,import,match,mixin,%
    new,null,object,override,package,% 
    private,protected,requires,return,sealed,% 
    super,this,trait,true,try,% 
    type,val,var,while,with,yield},%+
	otherkeywords={=,=>,<-,<\%,<:,>:,\#,@},%
   sensitive,%
   morecomment=[l]//,%
   morecomment=[n]{/*}{*/},%
   morestring=[b]",%
   morestring=[b]',%
   morestring=[b]""",%
  }[keywords,comments,strings]%

\lstdefinelanguage{jak}[]{Java}{
	morekeywords={refines,Super,layer}, 
	moredelim =*[is][\slshape\color{red}]{FF}{FF},%\underbar\colorbox{yellow}
	sensitive=f
}[keywords,comments,strings]

%% file: common.tex
\makeatletter

\newcommand{\side}[1]{}
\makeatletter 
\newcommand*{\bluntbreak}{\penalty\hyphenpenalty\hskip\z@skip} 

\makeatother

\newbox\subfigbox

\newcounter{fns}